\documentclass[aps, prd, preprint, showpacs, floatfix, superscriptaddress, nofootinbib, amsmath, amssymb]{revtex4-1}
\usepackage{amsfonts}
\usepackage{graphicx}
\usepackage{bm}
\usepackage{dcolumn}
\usepackage{color}
\usepackage{float}
\usepackage{relsize}
\usepackage{hyperref}
\usepackage{appendix}
\usepackage{slashed}
\begin{document}
\title{Decay constants of the pion and its excitations on the lattice.}

\author{Ekaterina V. Mastropas}
\affiliation{College of William and Mary, Williamsburg, VA 23187, USA}
\author{David G. Richards}
\affiliation{Jefferson Laboratory, 12000 Jefferson Avenue, Newport News, VA 23606, USA\\
			(the Hadron Spectrum Collaboration)} 

\begin{abstract}
  We present a calculation using lattice QCD of the ratios of decay
  constants of the excited states of the pion, to that of the pion
  ground state, at three values of the pion mass between 400
  and 700 MeV, using an anisotropic clover fermion action with three
  flavors of quarks.  We find that the decay constant of the first
  excitation, and more notably of the second, is suppressed with
  respect to that of the ground-state pion, but that the suppression
  shows little dependence on the quark mass.  The strong suppression
  of the decay constant of the second excited state is consistent with
  its interpretation as a predominantly hybrid state.
\end{abstract}
\maketitle

\section{Introduction}
Obtaining precise information about excited hadrons poses numerous
challenges. The principal computational challenge arises from the
faster decay of their Euclidean correlation functions in comparison
with those of the ground state, which leads to the worsening of the
signal-to-noise ratio. Additional complications arise in constructing
hadronic operators, where we seek to balance the computational cost
with the level of overlap achieved by a set of operators.

Despite these obstacles, advances in computational lattice QCD
techniques are such that precise quantitative calculations that can
confront both existing and forthcoming experiments are increasingly
feasible.  Experiments include those at the 12~GeV upgrade of the
Continuous Electron Beam Accelerator Facility (CEBAF) at Jefferson Lab
\cite{Dudek:2012vr}, with its new meson spectroscopy program in the
mass range up to 3.5~GeV. The expectation is that new data produced in
such experiments, combined with recent lattice QCD results aimed at
extracting the spectrum of the excited states for both mesons and
baryons \cite{Dudek:2009qf, Dudek:2010wm, Dudek:2011pd, Dudek:2011tt,
  Dudek:2012rd}, will represent a unique opportunity for the study of
the nature of confinement mechanism, and for identifying the role of
gluonic degrees of freedom in the spectrum.

The theoretical work presented here is devoted to the study of some of
the properties of excited states. It represents the first step in a
program to investigate quark distribution amplitudes, which can be
extracted from the vacuum-to-hadron matrix elements of quark bilinear
operators in the case of mesons, and of three-quark operators in the
case of baryons. These amplitudes can be used to provide predictions
for form factors and transition form factors at high momentum
transfers; in the case of baryons, the study of transition form
factors at high $Q^2$ is a focus of the JLab CLAS12 experiment, with
the aim of exploring the transition from hadronic to
quark-and-gluon-dominated dynamics.

In this paper, we provide details of a calculation of the leptonic
decay constant of the pion - the lightest system with a valence
quark-antiquark structure - and its excitations.  A knowledge of the
decay constants of the excited states, as well as of the ground state,
is important in delineating between different QCD-inspired pictures of
the meson spectrum, as well as demonstrating the feasibility of
studying the properties of highly excited states within lattice QCD.

The layout of the remainder of the paper is as follows.  In the next
section, we outline briefly the significance of the pseudoscalar decay
constants, and the state of our understanding for the pion. We then
describe our computational methodology for extracting the decay
constants not only of the ground-state pion, but of its excitations,
and provide details of the lattices used in our calculation.  In
Section~\ref{sec:results} we present our results, and compare to
expectations from models, and from previous lattice studies.  A
summary and conclusions are given in Section~\ref{sec:summary}, while
details of some derivations are provided in Appendix.

\section{Pseudoscalar leptonic decay constants}
Charged mesons are allowed to decay, through quark-antiquark
annihilations via a virtual $W$ boson, to a charged lepton and
(anti-)neutrino. The decay width for any pseudo-scalar meson $P$ of a
quark content $q_1\bar{q}_2$ with mass $m_P$ is given by
\begin{equation}
\Gamma (P \to l\nu)=\frac{G_f^2}{8\pi}f_P^2m_l^2m_P\left(1-\frac{m_l^2}{m_P^2}\right)|V_{q_1\bar{q}_2}|^2.
\label{eq:1}
\end{equation}
Here $m_l$ is the mass of the lepton $l$, $G_F$ is the Fermi coupling
constant, $V_{q_1\bar{q}_2}$ is the Cabibbo-Kobayashi-Moskawa (CKM)
matrix element between the constituent quarks in $P$, and $f_P$ is the
decay constant related to the wave-function overlap of the quark and
antiquark. A charged pion can decay as $\pi \to l \nu$ (we assume here
$\pi ^+\to l^+ \nu_l$ or $\pi^- \to l^-\bar{\nu}_l$), and its decay
constant $f_{\pi}$, which dictates the strength of these leptonic pion
decays, has a significance in many areas of modern physics. Thus a
knowledge of $f_{\pi}$ is important for the extraction of certain CKM
matrix elements, where the leptonic decay width $\Gamma$ in
Eqn. \eqref{eq:1} is proportional to $f_P|V_{q_1\bar{q}_2}|$.  The
pion decay constant, through its r\^{o}le in determining the strength
of $\pi \pi$ interactions, also serves as an expansion parameter in
Chiral Perturbation Theory \cite{Weinberg:1979, Gasser:1984}.  As
$|V_{ud}|$ has been quite accurately measured in super-allowed
$\beta$-decays, measurements of $\Gamma (\pi^+\to \mu^+\nu)$ yield a
value of $f_{\pi}$. According to PDG \cite{Beringer:2012pdg}, the most
precise value of $f_{\pi}$ is
\begin{equation}
f_{\pi^-}=(130.41\pm 0.03 \pm 0.20) \,\, {\rm MeV}.
\label{eq:2}
\end{equation}

Lattice QCD enables \textit{ab initio} computations of the mass
spectrum and decay constants of pseudo-scalar mesons, and the
calculation of the decay constant for ground-state mesons has
been an important endeavor in lattice calculations for the reasons
cited above. Recent lattice predictions
\cite{Follana:2008prl, Durr:2010hr, Bazavov:2010hj, Dowdall:2013rya}
for the ratio $f_K/f_{\pi}$ of $K^-$ and $\pi^-$ decay constants were
used in order to find a value for $|V_{us}|/|V_{ud}|$ which, together
with the precisely measured $|V_{ud}|$, provides an independent
measure of $|V_{us}|$.

The leptonic decay constant has a further role in hadronic physics in
representing the wave function at the origin, and therefore a
knowledge of the decay constant not only of the lowest-lying state but
of some of the excitations is important in confronting QCD-inspired
descriptions of the meson spectrum.  The pion excited states decay
predominantly through strong decays, and therefore experimental data
on their decay constants are lacking.  A study based on
Schwinger-Dyson equations \cite{Holl:2004fr} predicted significant
suppression of the excited-state pion decay constant in comparison
to that of the ground state. Similar
predictions, based on the QCD-inspired models and sum rules, also
propose remarkably small values for the decay constant of the first
pion excitation $f_{\pi_1}$; e.g., \cite{Volkov:1996br} proposed the
ratio $\frac{f_{\pi_1}}{f_{\pi_0}}$ to be of the order of one percent.
The authors of Ref.~\cite{Chang:2011vu} in their review of meson note
that the prediction in the chiral limit
\begin{equation}
f_{\pi_N}\equiv 0, N \geq 0 \nonumber
\end{equation}
is perhaps surprising, even though some some suppression of the
leptonic decay constants might be expected; for $S$-wave states, the
decay constant is proportional to the wave function at the origin, and
for excited states the configuration-space wave function is broader.
The only lattice study of the decay constants of the excited state of
the pion is that of the UKQCD Collaboration \cite{McNeile:2006qy},
where they obtained $f_{\pi_1}/f_{\pi_0}=0.078(93)$ in the chiral
limit, using an improved axial-vector current.  We will discuss these
results in further detail later.

\section{Computational Method}
The procedure for extracting energies and hadron-to-vacuum matrix
elements from a lattice calculation is to evaluate numerically
Euclidean correlation functions of operators
$\mathcal{O}_i$ and $\mathcal{O}_j$ of given quantum numbers, which are then
expressed through
their spectral representation
\begin{equation}
C_{ij}(t,0)=\frac{1}{V_3}\sum_{\vec{x},\vec{y}} \langle \mathcal{O}_i(\vec{x},t) \mathcal{O}^{\dagger}_j(\vec{y},0)\rangle
=\sum_N\frac{Z_i^{N*} Z_j^{N}}{2E_N} e^{- E_N t}.
\label{eq:3}
\end{equation}
Here $Z_N$ is the overlap of the $N^{\rm th}$ state in the spectrum, $\pi_N$,
\begin{equation}
  Z_i^N \equiv  \langle \pi_N \mid {\cal O}^{\dagger}_i(0) \mid 0 \rangle, \label{eq:4}
\end{equation}
$E_N$ is the energy of the state, and $V_3$ is the spatial
volume\footnote{Note that the correlation function defined here
  differs by a factor of $V_3$ from that of
  \protect\cite{Dudek:2010wm} so as to avoid an implicit factor of
  $\sqrt{V_3}$ in $Z_j^{N}$ and other matrix elements.}.  The ability to
  perform the time-sliced sum at both source and sink is a major
  benefit of the ``distillation" method used in our calculation.

  The representation in Eqn.~\eqref{eq:3} exposes some of the
  challenges in the study of hadronic excitations. The contributions
  of the excited states are suppressed exponentially, and the
  extraction of subleading exponentials is a demanding problem. As we
  climb up the spectrum, the signal-to-noise ratio tends to worsen
  with increasing $t$ (correlation functions decrease rapidly while
  statistical noise does not), and obtaining signals from the higher
  excitations becomes more and more problematic. Our means of
  overcoming these challenges is dependent on three novel
  elements. Firstly, the use of anisotropic lattices with a finer
  temporal than spatial resolution enabling the time-sliced
  correlators to be examined at small Euclidean times. Secondly, the
  use of the variational
  method~\cite{Michael:1985,Luscher:1990ck,Blossier:2009kd} with a
  large basis of operators derived from a continuum construction yet
  which satisfy the symmetries of the lattice.  Finally, an efficient
  means of computing the necessary correlation functions through the
  use of ``distillation"~\cite{Peardon:2009gh}.

\subsection{Gauge Configurations}
We employ the $N_f = 2 \oplus 1$ anisotropic lattices generated by the
Hadron Spectrum collaboration, with two mass-degenerate light quarks
of mass $m_l$ and a strange quark of mass $m_s$.  The lattices employ
improved gluon and ``clover" fermion actions, with stout smearing
restricted to the spatial directions.  Details are contained in
Refs.~\cite{Edwards:2008ja} and~\cite{Lin:2009rd}. Here we employ
$16^3\times 128$ lattices having a spatial lattice spacing of $a_s
\simeq 0.123~{\rm fm}$, and a renormalized anisotropy, the ratio of
the spatial and temporal lattice spacings, of $\xi \simeq 3.5$. The
calculations are performed at three values of the light-quark masses,
corresponding to pion masses of 391, 524 and 702 MeV. The 702 MeV pion
mass corresponds to the $SU(3)$-flavor-symmetric point.  The
parameters of the lattices used here are shown in
Table~\ref{tab:lattices}. The mass of the $\Omega$-baryon is used to
set the scale, and was determined within an estimated uncertainty of
2\% in Ref.~\cite{Bulava:2010yg} on the same ensembles; to facilitate
comparison with other calculations, we also provide the value of the
Sommer parameter $r_0$ on each ensemble.
\begin{table}
\centering
\begin{tabular}{cc|ccccc}
  $N_s$ & $N_t$ & $a_tm_l$ & $a_tm_s$ & $a_tm_{\pi}$ & $r_0/a_s$& $N_{\rm cfg}$  \\ \hline                                       
  16         &   128   &  -0.0743   & -0.0743       &   0.1483(2) & 3.21(1)    &    535         \\ 
  16         &   128   &  -0.0808   & -0.0743       &  0.0996(6)  & 3.51(1)    &    470        \\ 
  16         &   128   &  -0.0840   & -0.0743       &  0.0691(6)   & 3.65(1)   &    480        \\ 
\end{tabular}
\caption{Lattice extents ($N_s$ and $N_t$), the bare masses of light quark
  $a_tm_l$ and strange quark $a_tm_s$, the pion mass $a_tm_{\pi}$, the
  Sommer scale $r_0$, and the number $N_{\rm cfg}$ of gauge-field
  configurations.  On each configuration, solution vectors are
  computed from $N_{\rm vecs} = 64$ distillation
  vectors \protect\cite{Peardon:2009gh}, located on a single time slice.\label{tab:lattices} }
  \label{tab:1}
\end{table}

\subsection{Variational method}
A detailed description of the Hadron Spectrum Collaboration
implementation of the variational method can be found in
Ref.~\cite{Dudek:2010wm}, but we summarize it briefly here.  The
approach involves the solution of the generalized eigenvalue equation
\begin{equation}
C(t)v^{(N)}(t,t_0)=\lambda_N(t,t_0)C(t_0)v^{(N)}(t,t_0).
\label{eq:5}
\end{equation}
At sufficiently large $t > t_0$, the ordered
eigenvalues satisfy
\[
\lambda_N(t,t_0) \longrightarrow e^{ - E_N (t - t_0)},
\]
where $E_N$ is the energy of the $N^{\rm th}$ state. The eigenvalues
are normalized to unity at $t = t_0$, whilst the eigenvectors satisfy
the orthogonality condition:
\begin{equation}
v^{(N')\dagger}C(t_0)v^{(N)}=\delta_{N, N'}.
\label{eq:6}
\end{equation}
Identifying the energy of the $N^{\rm th}$ state with its mass, the
overlap factors $Z_i^N$ of the spectral representation are
straightforwardly related to the eigenvectors through
\begin{equation}
Z_i^N=\sqrt{2m_N}e^{m_Nt_0/2}v_j^{(N)*}C_{ji}(t_0).
\label{eq:7}
\end{equation}
We can define an ``ideal" operator
\begin{equation}
\Omega_N = \sqrt{2
  m_N} e^{-m_N t_0/2} v_i^{(N)} {\cal O}_i
  \label{eq:8}
\end{equation}
within the operator space for the state $N$ \cite{Dudek:2009kk}, where
the $v$'s are obtained from the solution of the generalized eigenvalue
equation at some $t= t_{\rm ref}$, and the operators are normalized so
as to remove the dependence on $t_0$.  

\subsection{Interpolating operator basis}
The efficacy of the variational method relies on an operator basis
that faithfully spans the low-lying spectrum.  The construction of
single-particle elements of such a basis is described in detail in
Refs.~\cite{Dudek:2009qf} and~\cite{Dudek:2010wm}.  Briefly, each
operator is constructed from elements of the general form
\begin{equation}
\bar{\psi}\Gamma\overleftrightarrow{D}_i\overleftrightarrow{D}_j\, ...\psi,
\label{eq:17}
\end{equation}
where $\overleftrightarrow{D}\equiv
\overleftarrow{D}-\overrightarrow{D}$ is a discretization of
gauge-covariant derivatives, and $\Gamma$ is one of the sixteen Dirac
matrices.  We then form an operator of definite $J$ and $M$, which we
denote by
\begin{equation}
\mathcal{O}^{J,M}=\left ( \Gamma\times D_{J_D}^{[N]}\right )^J.
\label{eq:18}
\end{equation}
We note that both charge conjugation, for neutral particles, and
parity are good symmetries on the lattice, but the full
three-dimensional rotational symmetry of the continuum is reduced to
the symmetry group of a cube. In the case of integer spin, there are
only five lattice irreducible representations, \emph{irreps}, labelled
by $\Lambda$ with row $\lambda$, instead of infinite number of
irreducible representations labelled by spin $J$ in the continuum. For
this study we are interested in mesons of spin $0$, lying in the $A_1$
\emph{irrep}; we note that this \emph{irrep} also contains continuum
states of spin $4$ and higher.  The subduction from the continuum
operators $\mathcal{O}^{J,M}$ of Eqn.~\eqref{eq:18} onto the lattice
\textit{irreps} denoted by $\Lambda$ and row $\lambda$ is performed
through the projection formula
\begin{equation}
\mathcal{O}^{[J]}_{\Lambda, \lambda}=\sum_MS^{\Lambda,\lambda}_{J,M}\mathcal{O}^{J,M},
\label{eq:19}
\end{equation}
where $S_{J,M}^{\Lambda\lambda}$ are the subduction coefficients. 

We use all possible continuum operators with up to three derivatives,
yielding a basis of 12 operators.  An important observation is that
for the ``single-particle" operators used here, there is remarkable
manifestation of continuum rotational symmetry at the hadronic scale,
that is the subduced operators of Eqn.~\eqref{eq:19} retain a memory
of their continuum antecedents \cite{Dudek:2009qf,Dudek:2010wm}.  One
of the operators arises from a continuum operator of spin 4.
Several operators, notably two of the form $\left(\Gamma \times
  D_{J=1}^{[2]}\right)^{J = 0}$, corresponding to the coupling of a
chromomagnetic gluon field to the quark and antiquark; these operators
are used as signatures for ``hybrid'' states with manifest gluonic
content.

The combination of the variational method, our operator constructions,
and the distillation method, described below, applied to the
anisotropic lattice ensembles has been shown to be very effective in
studies of excited light isovector mesons \cite{Dudek:2009qf,
  Dudek:2010wm}, isoscalar mesons \cite{Dudek:2011tt, Dudek:2013yja},
mesons containing charmed quarks \cite{Moir:2013ub,Liu:2012ze} and of
baryons
\cite{Edwards:2011jj,Dudek:2012ag,Edwards:2012fx,Padmanath:2013zfa}. We
now show how to exploit this toolkit to extract the vacuum-to-hadron
matrix elements of excited states.

\subsection{Axial-vector Current}\label{current}
The decay constant of the $N^{\rm th}$ excitation of the pion, $\pi_N$,
is given by the hadron-to-vacuum matrix matrix element of the axial
vector current,
\begin{equation}
\langle 0 \mid A_{\mu}(0) | \pi_N \rangle = p_{\mu} f_{\pi_N},
\label{eq:10}
\end{equation}
where $A_{\mu } = \bar{\psi} \gamma_{\mu} \gamma_5 \psi$; for a state
at rest, considered here, only the temporal component of the matrix
element is non-zero.  The matrix element of the axial-vector current
determined on an \textit{isotropic} lattice is related to that in
some specified continuum renormalization scheme through an operator matching
coefficient $Z_A$:
\begin{equation}
A_{\mu} = Z_A A_{\mu}^{\rm lat}. \label{eq:ZA}
\end{equation}
$Z_A$ is unity to tree level in perturbation theory, and furthermore
the mixing with higher-dimension operators at ${\cal O}(a)$ only
occurs at one-loop.  However, on an anisotropic lattice, mixing with
higher dimension operators occurs at
tree level \cite{Chen:2001hq}.  For the action employed here, we
find:
\begin{equation}
A_4^{\rm I}  = ( 1 + m a_t \Omega_m) \left[ A_4^{\rm U} - \frac{1}{4}(\xi - 1)
a_t \partial_4 P \right]\\ \label{eq:improv4}
\end{equation}
where $A_4^{\rm U} \equiv \bar{\psi} \gamma_4\gamma_5 \psi$ is
the temporal component of the unimproved local axial-vector current introduced earlier, and $P =
\bar{\psi} \gamma_5 \psi$ is the pseudoscalar current; the derivation
is provided in the Appendix.  There is an ambiguity in the values of
the parameters $m, \Omega_m, \xi$ at tree level, and in this work we
take $\xi$ to have its target renormalized value of 3.5.  It is
important to note that the mixing at tree level vanishes for an
isotropic action, $\xi = 1$, and therefore is an artefact of the
anisotropic action used in this work.  In our subsequent analysis, we
will consider the ratios of the decay constant of an excited state and
that of the ground state; both the matching coefficient of $Z_A$ of
Eqn.~\eqref{eq:ZA} and the mass improvement term $(1 + m a_t
\Omega_m)$ of Eqn.~\eqref{eq:improv4} cancel in these ratios.
Finally, to obtain the physical value of the decay constant from the
lattice value, we have \cite{Dudek:2006ej}
\begin{equation}
f_{\pi_N} = \xi^{-3/2} a_t^{-1} \tilde{f}_{\pi_N},
\end{equation}
where $\tilde{f}_{\pi_N}$ is the dimensionless value obtained in
our calculation.

Armed with the optimal interpolating operator for the $N^{\rm th}$
excited state, we now extract its lattice decay constant
$\tilde{f}_{\pi_N}$ through the two-point correlation function
\begin{equation}
  C_{A_4,N}(t) = \frac{1}{V_3} \sum_{\vec{x},\vec{y}} \langle 0 \mid A_4(\vec{x},t) \Omega^{\dagger}_N(\vec{y},0) \mid 0 \rangle
  \longrightarrow e^{- m_N t} m_N \tilde{f}_{\pi_N},
\label{eq:11}
\end{equation}
where $A_4$ is the temporal component of either the unimproved or
improved axial-vector current. Finally, we note that whilst the sign
of the decay constants has been discussed in Refs.~\cite{Qin:2011xq}
and \cite{Holl:2005vu}, the matrix element $\langle 0 \mid A_\mu \mid
\pi_N \rangle$ for both the improved and unimproved currents, obtained
through Eqn.~(\ref{eq:11}), is defined only up to a phase, since the
corresponding eigenvector $v^{(N)}$ can be multiplied by an arbitrary
phase.  We therefore quote the absolute values of the decay constants
in our subsequent analyses.

\subsection {Distillation}
Physically relevant signals in correlation functions fall
exponentially and are dominated by statistical fluctuations at
increasing times.  Therefore, it is essential to use  operators
with strong overlaps onto the low-lying states, and whose overlaps to
the high-energy modes are suppressed. If the interpolating
operators are constructed directly from the local fields in the
lattice Lagrangian, then the coupling to the high energy modes is
strong.  A widely adopted means of suppressing this coupling is
through the use of spatially extended, or ``smeared", quark fields.
We accomplish this smearing through the adoption of
``distillation"~\cite{Peardon:2009gh}, in which the distillation
operator has the following form:
\begin{equation}
\square_{x,y}(t)=\sum_{k=1}^{N_{\rm vecs}}\xi_x^{(k)}(t)\xi_y^{(k)\dagger}(t).
\label{eq:12}
\end{equation}
Here $\xi^{(k)}\, (k = 1,\dots,N_{\rm vecs})$ are the $N_{\rm vecs}$
eigenvectors of the gauge-covariant lattice Laplacian, $-\nabla^2$,
corresponding to the $N_{\rm vecs}$ lowest eigenvalues, evaluated on
the background of the spatial gauge-fields of time slice $t$.  A meson
interpolating operator then has the general form
\begin{equation}
{\cal O} =\bar{\tilde{\psi}}(t)\boldsymbol{\Gamma}\tilde{\psi}(t),
\label{eq:13}
\end{equation}
where $\tilde{\psi} = \square \psi$, $\boldsymbol\Gamma$ is an operator
acting in $\{position,\,spin,\,color\}$ space, and a correlation
function between operators ${\cal O}_i$ and ${\cal O}_j$ can be
written as
\begin{equation}
C_{ij}(t)=\langle \bar{\psi}(t)\square(t)\boldsymbol{\Gamma}^i(t)\square(t)\psi(t) \cdot \bar{\psi}(0)\square(0)\boldsymbol{\Gamma}^j(0)\square(0)\psi(0) \rangle.
\label{eq:14}
\end{equation}
Due to the small rank of its smearing operator, distillation has major
benefit over other smearing techniques in significantly reducing the
computational cost related to the construction of all elements of the
correlation matrix, whilst enabling a time-sliced sum to be performed
both at the sink and at the source.

The construction of the correlation functions from operators smeared
both at the sink and the source has been described in detail in
Ref. \cite{Peardon:2009gh}, but the extension to the calculation of
the smeared-local two-point functions needed here is straightforward.
Our starting point is the solution of the Dirac equation from the the
eigenvectors at time slice $t'$, which without loss of generality we
take to be on time slice $t' = 0$
\begin{equation}
\tilde{\tau}^{(k)}_{\alpha\beta}(\vec{x},t;t'=0) =
M^{-1}_{\alpha\beta}(\vec{x},t;t'=0) \xi^{(k)}(t'=0).
\label{eq:15}
\end{equation}
We then construct
\begin{eqnarray}
  C_{\mu,i}(t, 0) & = & 
\frac{1}{V_3} 
\sum_{\vec{x},\vec{y}}
  \langle 0 \mid A_{\mu} (\vec{x},t) {\cal O}^{\dagger}_i(\vec{y},0) \mid 0 \rangle \nonumber \\
  & = & 
\sum_{\vec{x}} {\rm Tr} [ 
  \gamma_\mu \tilde{\tau}(\vec{x},t; 0)  \Phi_i(0) \gamma_5 \tilde{\tau}(\vec{x}, t;0)^{\dagger}],
  \label{eq:16}
\end{eqnarray}
where the trace is over spin, color and eigenvector indices, and $\Phi$ is
the representation of the operator ${\cal O}_i$ in terms of the
eigenvectors $\xi$. The
correlator onto the optimal operator for the $N^{\rm th}$ excited
state immediately follows from Eqn.~\eqref{eq:11}.

\section{Results}\label{sec:results}
The determination of the excited-state spectrum using the variational
method has been described in detail in
Refs.~\cite{Dudek:2010wm,Dudek:2009qf}, and we merely present the
results for the spectrum of the lowest lying states as the first row
for each ensemble in Table~\ref{tab:5}; we quote only the lowest-lying
four states in the spectrum, since the next state is identified as
having spin 4, as we discuss later. In practice, the coefficients
giving rise to the ``optimal" operator for the $N^{\rm th}$ excited
states must be determined at some value $t_{\rm ref} > t_0$; we take
the value of $t_{\rm ref}$ as that which gives the best reconstruction
of the correlation matrix used in the variational method, following
the technique described in Ref.~\cite{Dudek:2010wm}.  The decay
constants $f_{\pi_N}$ are obtained through the correlation function
$C_{A_4,N}(t)$ of Eqn.~\eqref{eq:11}, using the optimal operator
determined above.  The mass spectrum obtained from a two-exponential
fits to these correlators, using the unimproved axial-vector current
at the sink, is listed in the second row for each ensemble in
Table~\ref{tab:5}.  The consistency between the resultant spectra is
encouraging.

\begin {table}%
\centering \small
\begin{tabular}{c|cccc}
$m_{\pi}~{\rm (MeV)}$ &  $N=0$ & $N=1$ & $N=2$ & $N=3$ \\ \hline  \hline                                    
702   & 0.1483(1)   &  0.3619(11)  & 0.4439(34)      &   0.5199(61) \\ 
                  & 0.1482(4)    & 0.3600(84)  & 0.3664(975)   &  0.5569(506)  \\ \hline
524 & 0.0999(5) & 0.3118(31)    & 0.4028(43)  & 0.4493(149)       \\ 
                  &  0.1008(4)  & 0.3134(99)   & 0.4047(683)   & 0.4361(460)      \\ \hline
391 & 0.0694(2) & 0.2735(31)  & 0.3665(34) & 0.4209(99)  \\
                  & 0.0709(10) & 0.2626(93)  & 0.3592(688)  & 0.4270(75) 
\end{tabular}
\caption{The first line for each ensemble lists the masses of the pion and its first
  three excitations in lattice units obtained from the variational
  method.  The second line
  lists the masses obtained from a two-exponential fit to the
  correlator of Eqn.~\protect{\eqref{eq:11}} using the optimal
  interpolating operator from the variational method at the source, and the unimproved axial-vector current at the sink.\label{tab:5}}
\end{table}

\begin{figure}[t]
\begin{center}
  \includegraphics[width=0.45
  \textwidth]{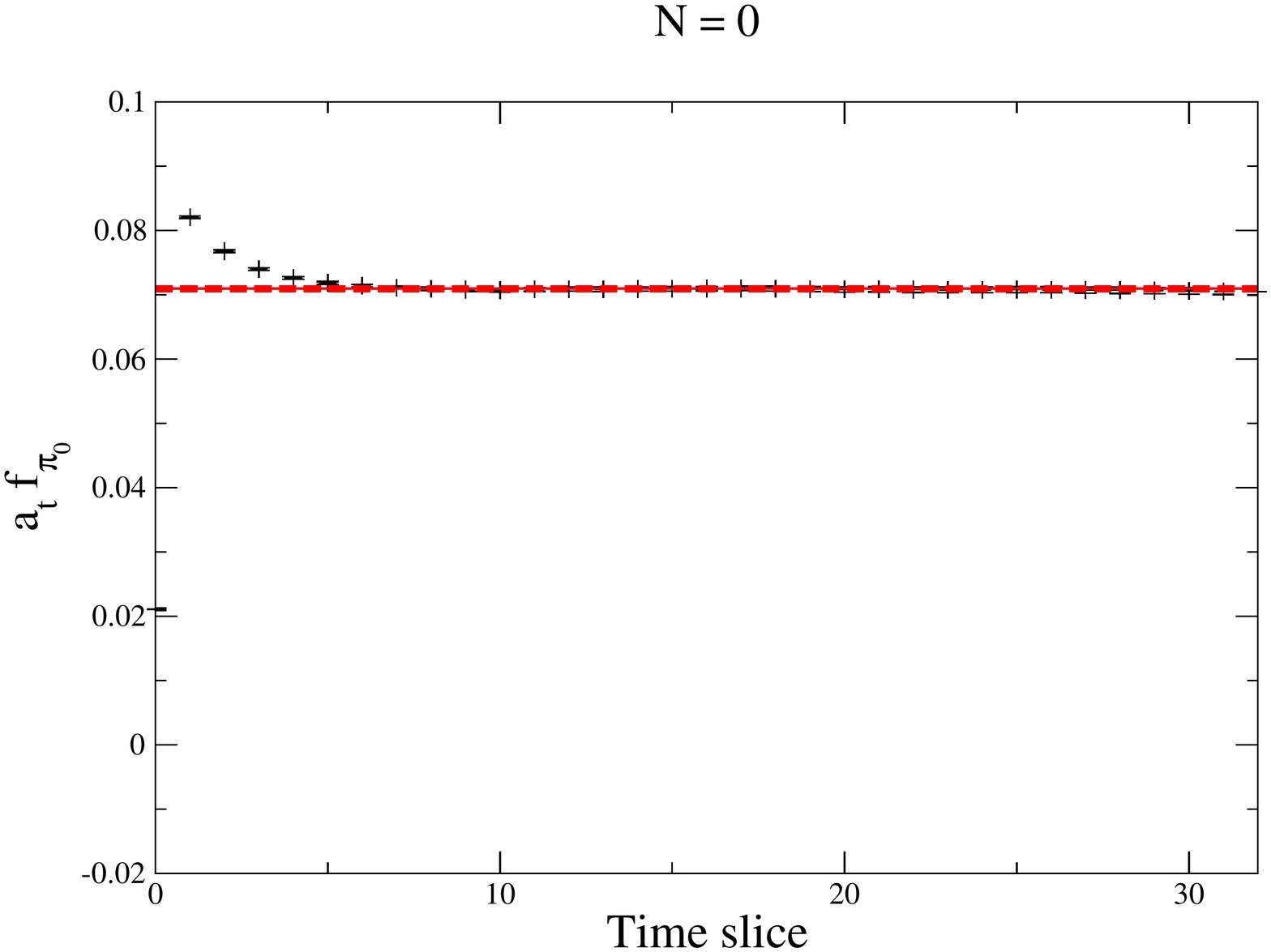}\hfill\includegraphics[width=0.45\textwidth]{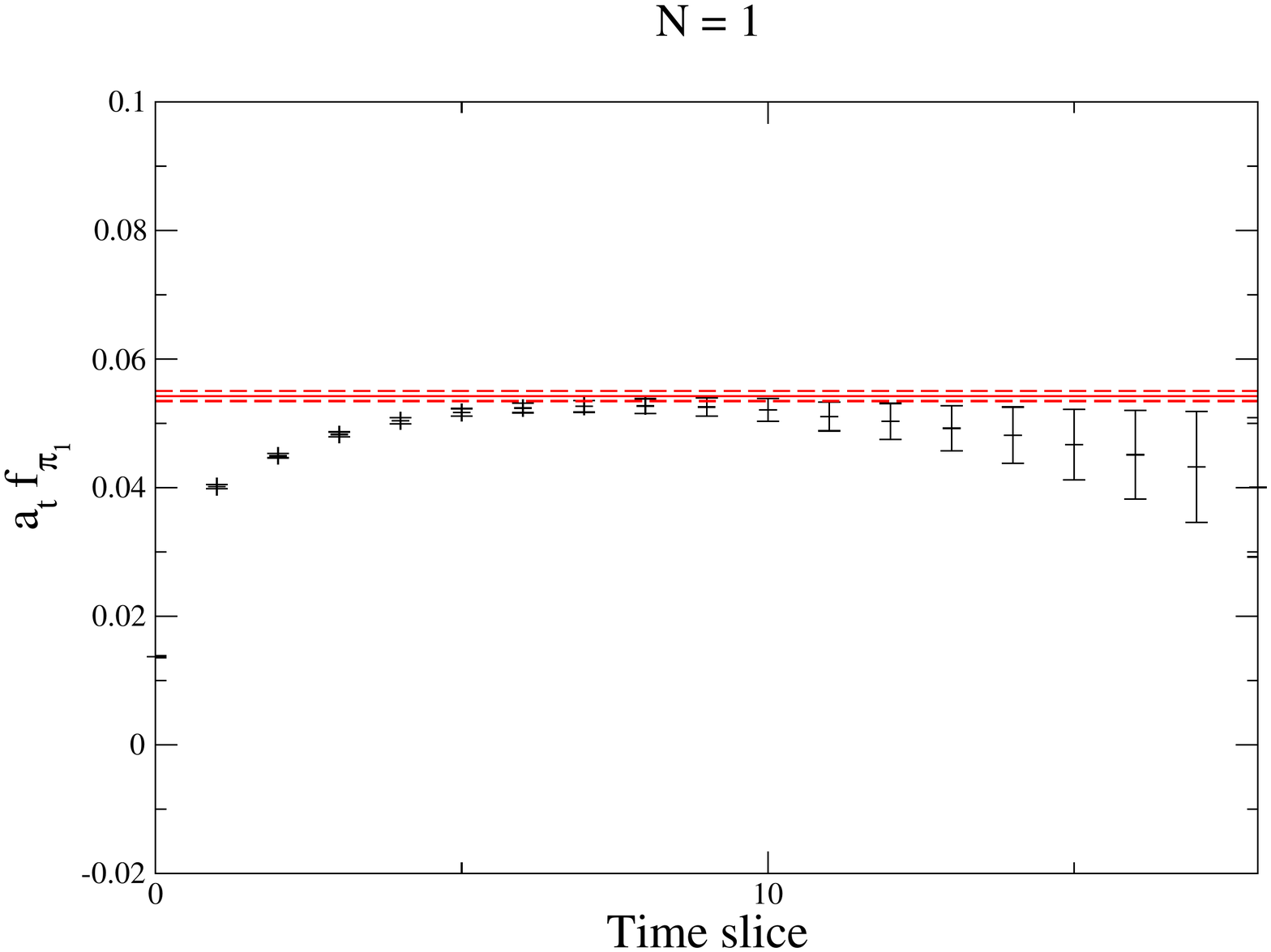}\\
  \includegraphics[width=0.45
  \textwidth]{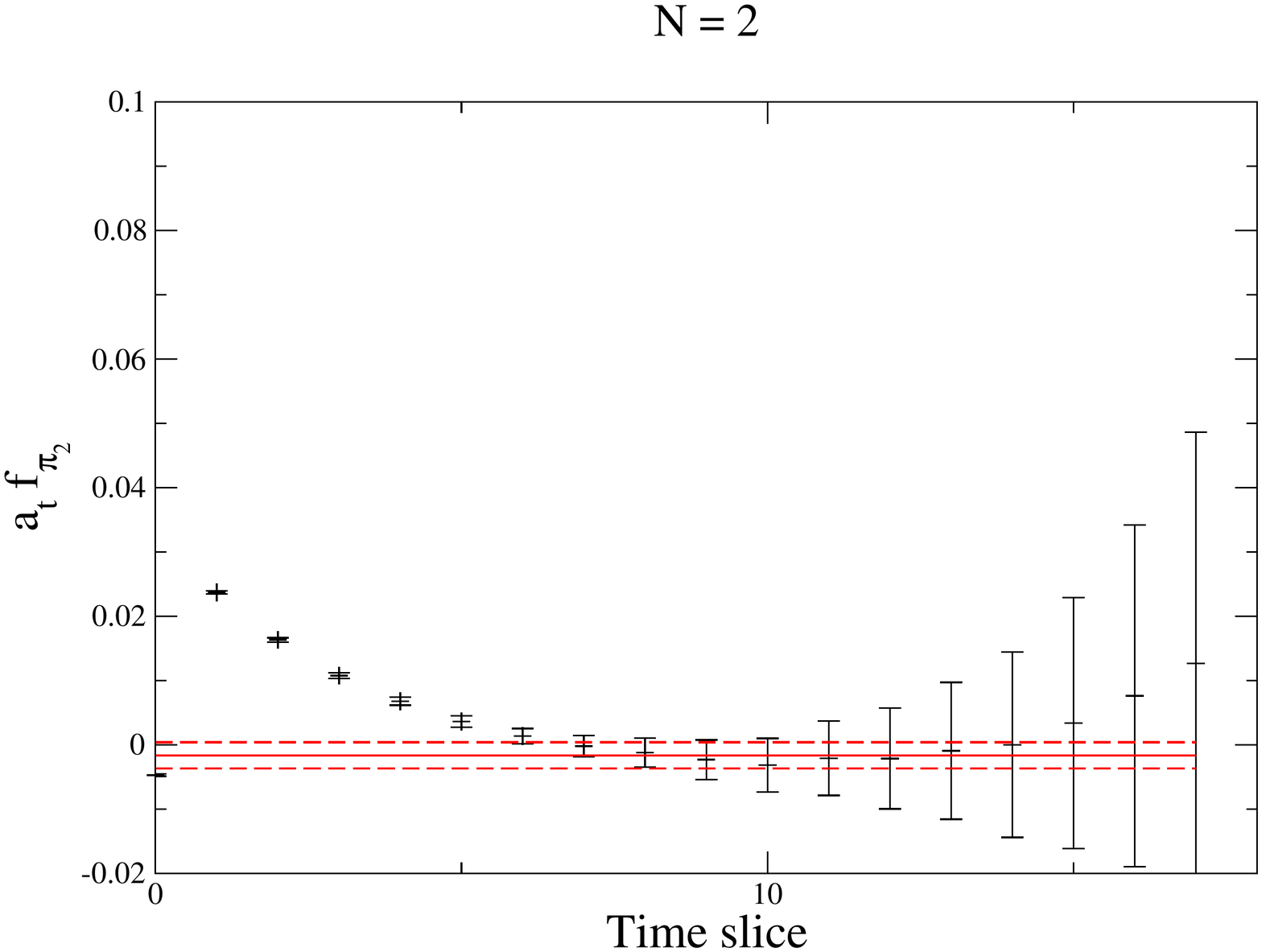}\hfill\includegraphics[width=0.45\textwidth]{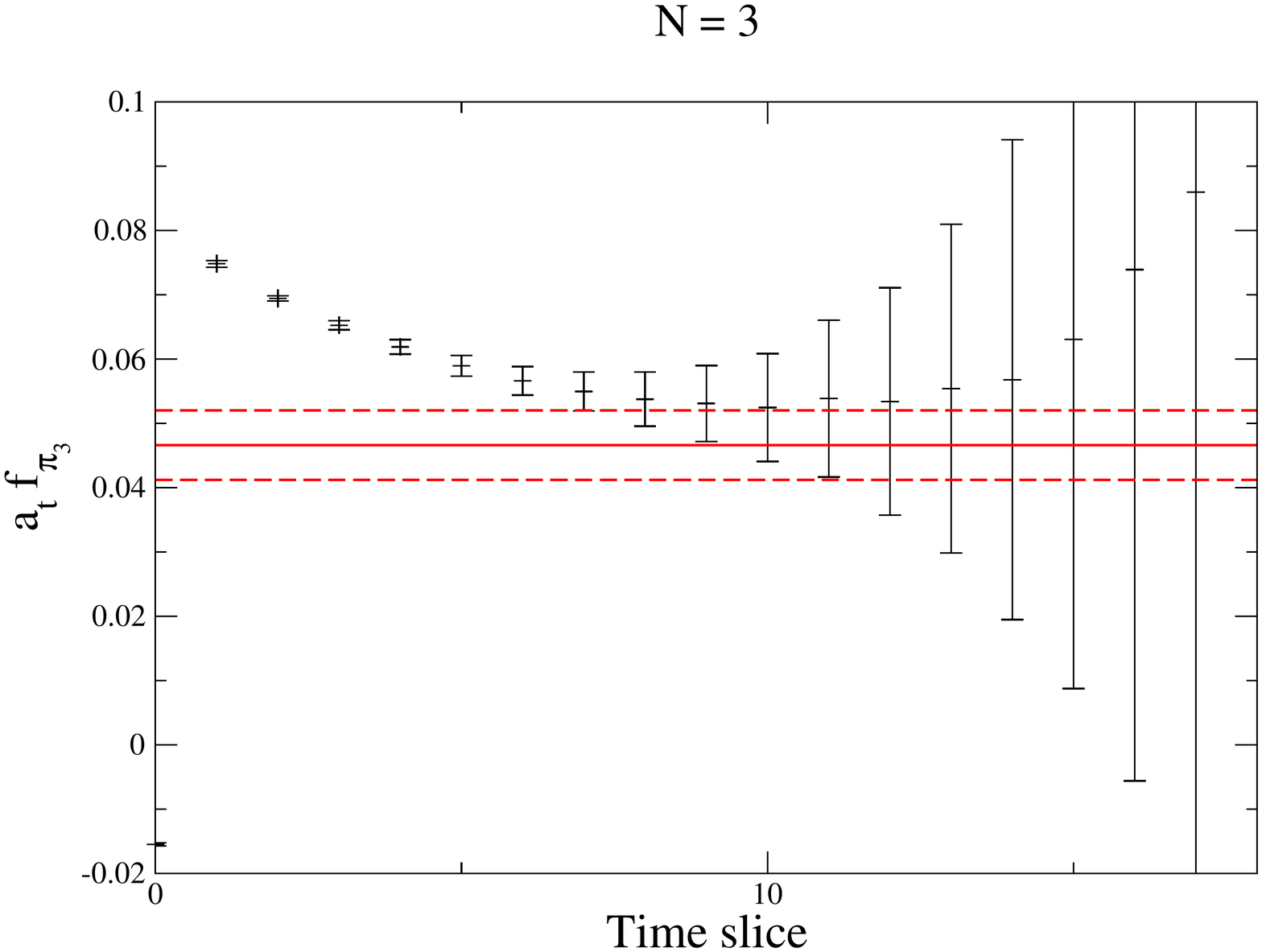}
\end{center}
\caption{The data for $a_t f_{\pi_N}$ in units of the temporal lattice
  spacing from Eqn.~\eqref{eq:comb}, for the ensemble at $m_\pi =
  702~{\rm MeV}$; the line corresponds to the value of $a_t f_{\pi_N}$
  obtained from a three-parameter fit to the data as discussed in the
  text.  The optimal operators are obtained from the variational
  method with $t_0 = 7$ and the eigenvectors determined at $t_{\rm
    ref} = 15$. \label{fig:fpi_fit}}
\end{figure}

\begin {table}
\centering \small
\begin{tabular}{c|cccc}
  $m_{\pi}~{\rm (MeV)}$ &  $N=0$ & $N=1$ & $N=2$ & $N=3$ \\ \hline  \hline   
  702 & 0.0551(3)   &  0.0319(10)  & 0.0005(12)      &   0.0307(23) \\                                  
  & 0.0716(6)    & 0.0556(52)  & 0.0041(23)   &  0.0565(54)\\ 
  & 0.0710(4)    &  0.0543(8)  & 0.0017(21)   &  0.0466(54)\\ \hline
  524 & 0.0441(5) & 0.0261(12)    &0.0057(3)  &  0.0315(31)       \\ 
  & 0.0565(18)  & 0.0465(27)   & 0.0065(43)   &  0.0493(132)    \\ 
  & 0.0564(6)  & 0.0476(62)   & 0.0083(10)   & 0.0483(91)      \\ \hline
  391 &  0.0369(7) & 0.0218(15)  & 0.0062(18) & 0.0256(5)      \\
  &  0.0476(8)  & 0.0429(113)   & 0.0138(28)  & 0.0508(11) \\
  & 0.0473(9)  & 0.0398(90)   & 0.0140(67)  & 0.0462(11) 
\end{tabular}
\caption{The unrenormalized values of $a_t f_{\pi_N}$ for the ground
  state and first three excitations. For each ensemble, the first line
  are the values computed using the unimproved axial-vector current,
  while the second and third lines employ the improved axial-vector
  current of Eqn.~\eqref{eq:improv4} with the derivative of the
  pseudoscalar current computed using the corresponding energy of the
  state, and a finite time difference, respectively.}
 \label{tab:imp}
\end{table} 
In order to extract the matrix element, we form the combination
\begin{equation}
e^{m_Nt} C_{A_4,N}(t)/m_N \longrightarrow \tilde{f}_{\pi_N} + B_N e^{- \Delta
  m_N t}, \label{eq:comb}
\end{equation}
using the mass $m_N$ obtained through the variational method.  A
three-parameter fit in $\{\tilde{f}_{\pi_N}, B_N, \Delta m_N\}$ then
yields the value of the decay constant.  In Table~\ref{tab:imp} we
present, as the first line for each ensemble, our results for the
absolute, unrenormalized values of the pion decay constants $a_t
f_{\pi_N}$ for the ground ($N=0$) and first three excited states
($N=1,\,2\,,3$), obtained using the unimproved axial-vector current.
As discussed earlier, the use of an anisotropic lattice introduces
mixing with higher dimension operators, even at tree level.  We thus
calculate the decay constants through Eqn.~\eqref{eq:11}, but using
the improved axial-vector current of Eqn.~\eqref{eq:improv4}.  We can
evaluate the partial derivative of the pseudoscalar current contributing
to the improved current in two ways: by replacing it with energy of
the state, $ \partial_4 P \rightarrow E_N P$, and through the use of a
finite difference between successive time slices, $\partial_4 P
\rightarrow P(t+1) - P(t)$. These are presented as the second and
third rows for each ensemble in Table~\ref{tab:imp}. The two methods
of computing the temporal derivative are in general consistent, and we
will use the finite-difference method in the subsequent
discussion. Finally, as an illustration of the quality of our
procedure, we show in Figure~\ref{fig:fpi_fit} the data for
Eqn.~\eqref{eq:comb}, together with the values of $a_t f_{\pi_N}$ obtained
from the three-parameter fit, for the $N_f = 3$ ensemble.

\begin{figure}
\centering
\includegraphics[width=0.6 \textwidth]{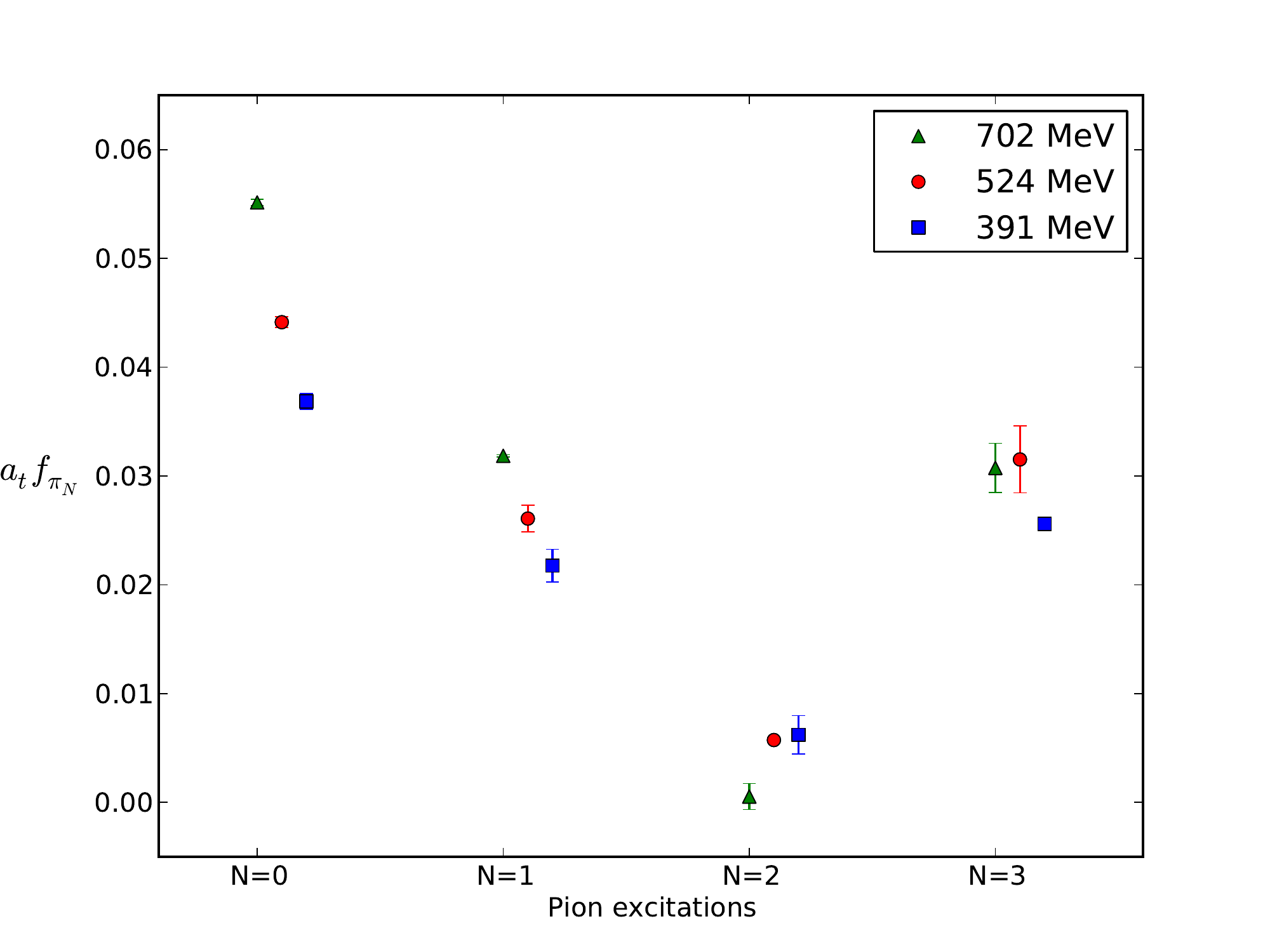}
\caption{The unrenormalized pion decay constants $a_t f_{\pi_N}$ on each
  of our ensembles 
  obtained using the unimproved axial-vector current.}
\label{fig:1}
\end{figure} 
\begin{figure}
\centering
\includegraphics[width=0.6 \textwidth]{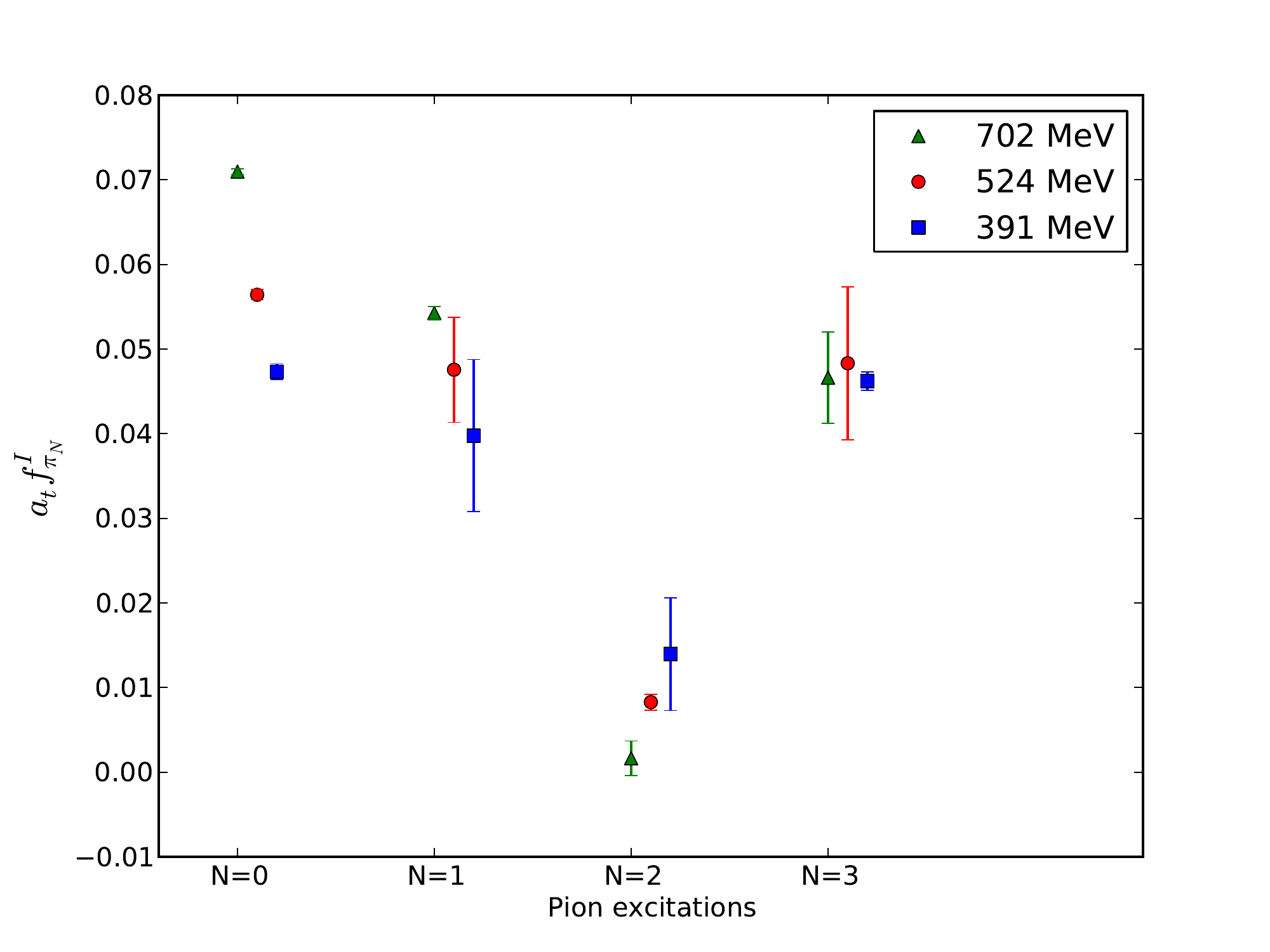}
\caption{The unrenormalized pion decay constants $a_t f_{\pi_N}$ on
  each of our ensembles
  obtained using the improved axial-vector current.}
\label{fig:improved_trend}
\end{figure}
\begin{figure}
\centering
\includegraphics[width=0.6 \textwidth]{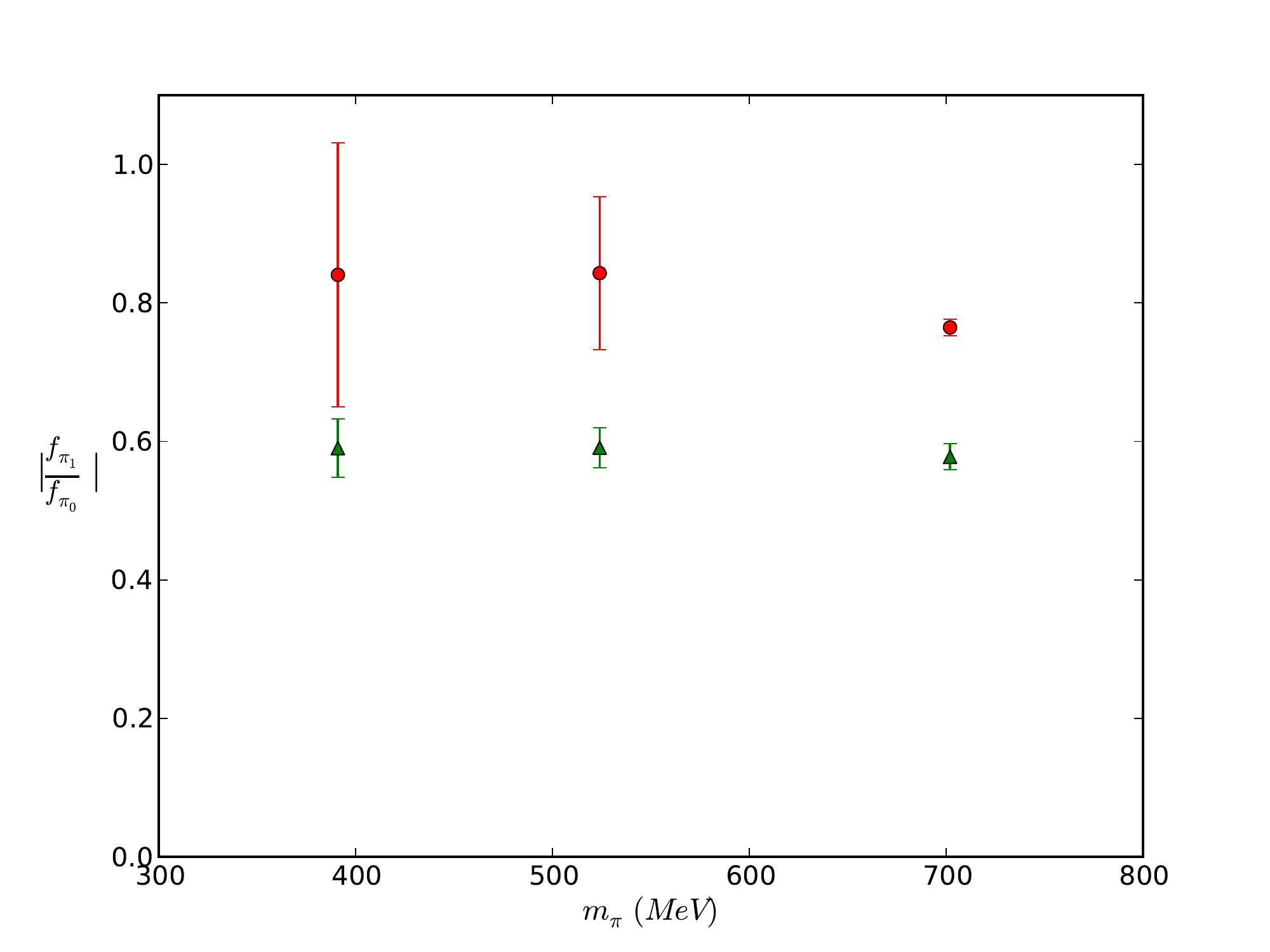}
\caption{Lattice values for the ratio of the ``improved" decay
  constants for the first excited $f_{\pi_1}$ and ground-state
  $f_{\pi_0}$ pion as a function of the pion mass. The green points
  represent unimproved values, while data in red color correspond to
  the ratios of improved decay constants.\label{tab:improved}}
\label{fig:plot_improved}
\end{figure}

 
The decay constants $a_t f_{\pi_N}$ for each of our ensembles computed
using the unimproved and improved axial-vector currents is presented
in Figures~\ref{fig:1} and~\ref{fig:improved_trend}, respectively.  We
observe a decrease in the value of the decay constant up to and
including that for the second excited state on all three ensembles,
irrespective of the use of the unimproved or improved axial-vector
current.  In Figure~\ref{fig:plot_improved}, we show the ratio of the
decay constant of the first excited state to that of the ground state,
a combination in which the matching factor cancels, for both the
unimproved (green) and improved (red) currents.  Whilst we note that
the improvement term represents a significant contribution at each
quark mass, once again the behavior of the ratios remains the same for
both currents.

\begin{figure}
\centering
\includegraphics[width=0.6 \textwidth]{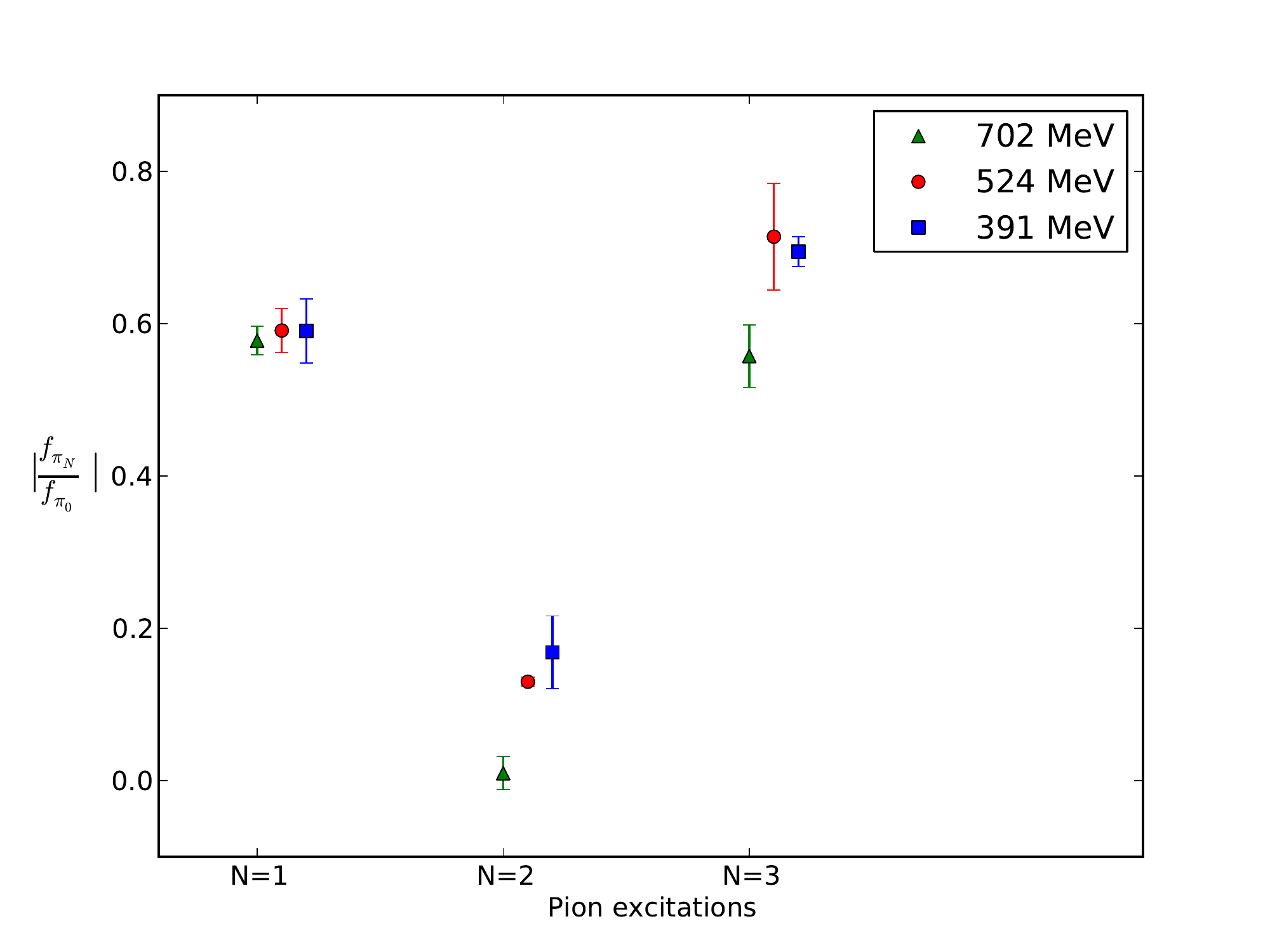}
\caption{Ratios of the excited-state decay constants $f_{\pi_N}$ to
  the ground-state decay constant $f_{\pi_0}$ for the first three pion
  excitations ($N=1,\, 2, \, 3$), using the unimproved current.}
\label{fig:3}
\end{figure}

\begin{figure}
\centering
\includegraphics[width=0.6 \textwidth]{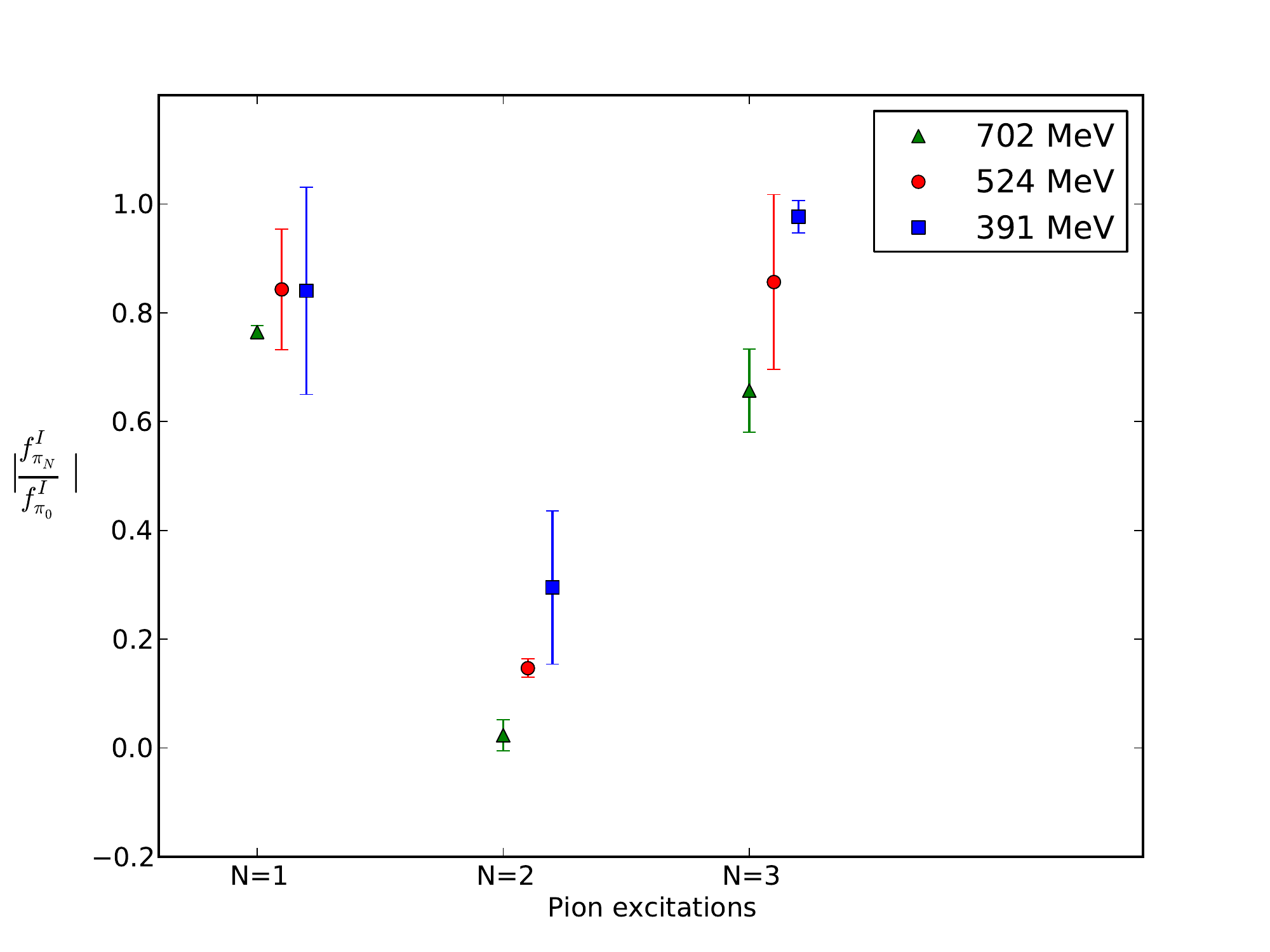}
\caption{Ratios of the excited-state decay constants $f_{\pi_N}$ to
  the ground-state decay constant $f_{\pi_0}$ for the first 3 pion
  excitations ($N=1,\, 2, \, 3$), using the improved current.}
\label{fig:improved_ratios}
\end{figure}

So far, all lattice QCD predictions for the decay constant of the
excitations of the pion have been made for the first excited state
only.  Here, we extend previous work through the calculation of the
decay constant of higher excitations, up to that of the third excited
state. The ratios $f_{\pi_N}/f_{\pi_0}$ of decay constants for the
1$^{\rm st}$, 2$^{\rm nd}$ and 3$^{\rm rd}$ excited states to that of
the ground state $f_{\pi_0}$ are shown using the unimproved and
improved currents respectively in Figures~\ref{fig:3}
and~\ref{fig:improved_ratios}, respectively.

Our results indicate the value of $\frac{f_{\pi1}}{f_{\pi0}}$ to be
largely independent of the pion mass in the explored region of
$400-700$ MeV. These conclusions differ from the previously mentioned
lattice study performed by UKQCD Collaboration \cite{McNeile:2006qy}.
They find in particular that their results show a strong dependence on
the current used.  A simple linear fit to the ratio of the improved
decay constants obtained through the implementation of the full ALPHA
Collaboration method \cite{Jansen:1995ck} gave
$|f_{\pi_1}/f_{\pi_0}|=0.078(93)$ in the chiral limit, showing a significant
suppression of the decay constant for the first pion
excitation. Meanwhile, for the unimproved decay constants, they
obtained $|f_{\pi_1}/f_{\pi_0}|=0.38(11)$ in the chiral limit.  We
have also employed an improved current, but the improvement term we
include arises at tree level and is an artefact of the use of an
anisotropic lattice.

\begin{figure}
\centering
\includegraphics[width=0.7\textwidth,angle=-90]{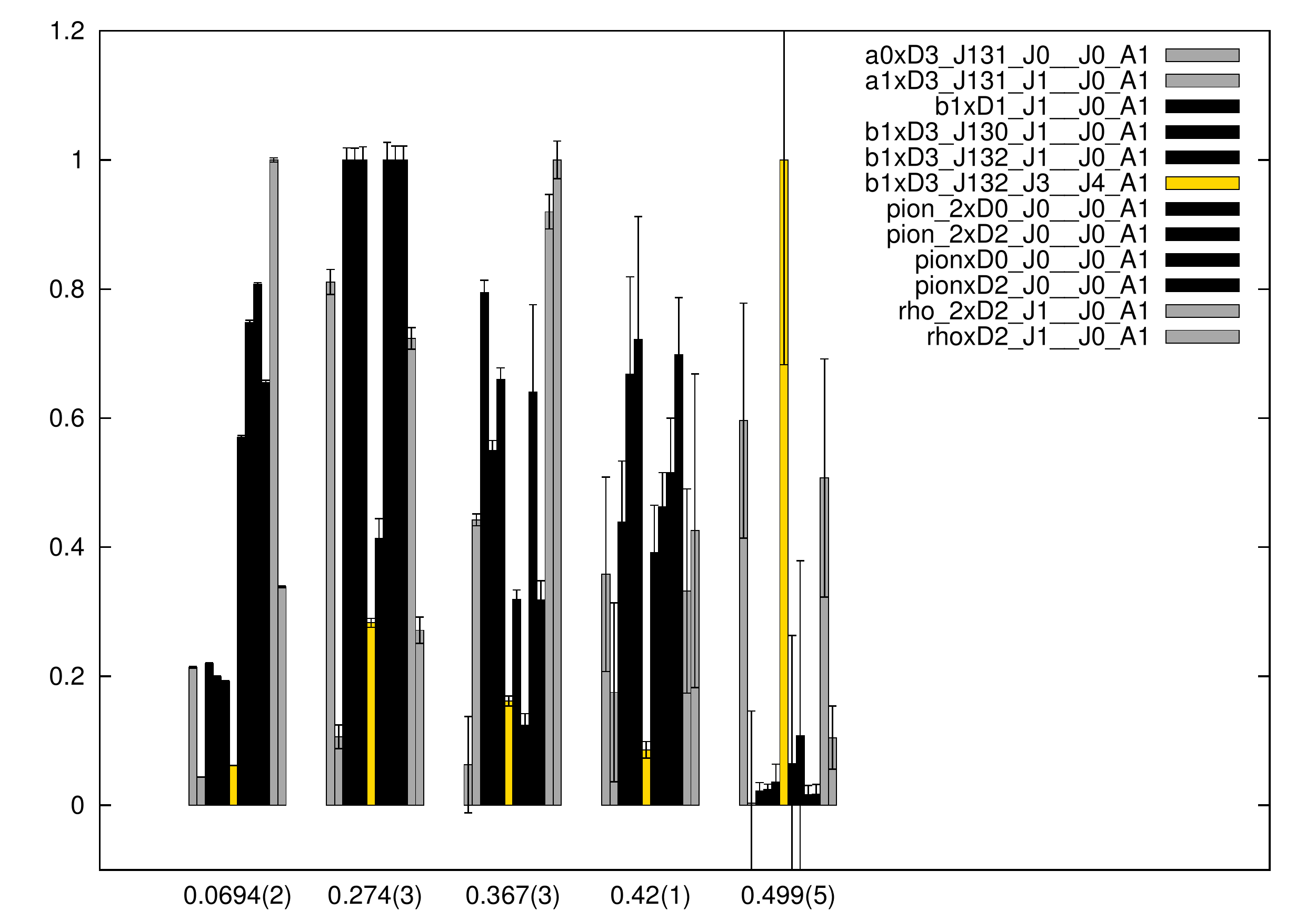}
\caption{The histogram shows the overlap of the operators of the
  variational basis to the five lowest-lying states in the spectrum,
  for the data corresponding to a pion mass of $391~{\rm MeV}$, as
  described in Refs.~\protect\cite{Dudek:2010wm,Dudek:2009qf}. The
  yellow bar denotes the overlap onto an operator derived from a $J=4$
  continuum construction; we associate the fourth excitation with a
  state of spin 4, and do not discuss further.  Grey bars denote
  overlaps onto ``hybrid'' operators, as discussed in the
  text.\label{fig:overlap840}}
\end{figure}

A particularly striking observation is the strong suppression of the
decay constant of the second excitation.  The quark and gluon content
of the excitations of the pion spectrum has been investigated earlier
using the overlaps of the operators of the variational basis with the
states in the spectrum as signatures for their partonic
content \cite{Dudek:2010wm, Dudek:2009qf}, and a phenomenological
interpretation provided in Ref.~\cite{Dudek:2011pd}.  Of the lowest
four states in the spectrum that we study here, each was identified as
corresponding to a state of spin $0$ rather than of spin $4$, with the
first excitation an $S$-wave radial excitation, but with the second
excited state having a significant hybrid content represented by a
strong overlap onto operators comprising a quark and antiquark coupled
to a chromomagnetic field, as we illustrate for the lightest ensemble
in Figure~\ref{fig:overlap840}.  Thus the strong suppression of the
decay constant for the second-excited state, but the far more moderate
suppression of the first excited state, is quite understandable within
this phenomenology.

\section{Conclusions}\label{sec:summary}
In this work, we have undertaken the first steps in investigating the
properties of the excited meson states in QCD by computing the decay
constants both of the pion, and of its lowest three excitations.  Our
results show that the optimal operators obtained through the
variational method are effective interpolating operators when
calculating the hadron-to-vacuum matrix elements of local operators.
The picture that emerges is that for the lowest two excitations, the
decay constants are indeed suppressed, but largely independent of the
quark mass, and that the strong suppression for the second excited
state is indicative of the predominantly hybrid nature of the state.

The work presented here is highly encouraging, but there are certain
caveats.  Firstly, the basis of interpolating operators used here
includes only ``single-hadron" operators, whose coupling to
multi-hadron decay states is expected to be suppressed by the volume,
and thus our results effectively ignore that higher excitations become
unstable under the strong interactions.  Our previous work on the
isovector spectrum suggested that the single-particle energy levels at
these values of the quark mass are somewhat insensitive to the volume,
but that has not been checked for the decay matrix elements.
None-the-less, the fact that the decay constant ratios themselves show
a limited quark-mass dependence, despite large differences in $m_\pi
L$ ($L$ being the length of the lattice), leads credence to the results presented here.  Secondly, the
improvement term we include in the axial vector current is that
arising at tree level through the use of an anisotropic action;
mixings beyond tree level, and the matching coefficients, which cancel
in the ratios of decays constants, have not been included.  As well as
addressing these issues, future work will extend the calculation to
obtain the moments of the quark distribution amplitudes, and will
investigate the decay constants and distribution amplitudes for both
the $\rho$ and nucleon excitations.

\section{Acknowledgements}
We thank our colleagues within the Hadron Spectrum Collaboration, and
in particular, Jo Dudek, Robert Edwards, Christian Shultz and
Christopher Thomas.  We are grateful for discussions with Zak Brown
and Hannes L.L.\ Roberts, who was involved at an earlier stage of this
work.  {\tt Chroma}~\cite{Edwards:2004sx} was used to perform this
work on clusters at Jefferson Laboratory under the USQCD Initiative
and the LQCD ARRA project.  We acknowledge support from
U.S. Department of Energy contract DE-AC05-06OR23177, under which
Jefferson Science Associates, LLC, manages and operates Jefferson
Laboratory.

\begin{appendix} 
\section{Axial-vector current improvement}
Here we provide derivation of the formula for the improved
axial-vector current that we use in our computations. Following
closely the discussion on the classical improvement of the anisotropic
action introduced in Ref. \cite{Chen:2001hq}, we first start with the
naive fermion action that has manifestly no $\mathcal{O}(a)$
discretization errors
\begin{equation}
\bar{\psi}_c(m_c+\slashed{\nabla})\psi_c \nonumber;
\label{}
\end{equation}
the bare quark mass $m_c$ here is the same as in the continuum. The
$\mathcal{O}(a)$-improved anisotropic quark action can be derived by
applying the field redefinition
$\bar{\psi}_c=\bar{\psi}\,\bar{\Omega}$ ($\psi_c=\psi\,\Omega$), where
\begin{eqnarray}
&& \Omega=1+\frac{\Omega_m}{2}a_tm_c+\frac{\Omega_t}{2}a_t\overrightarrow{\slashed{\nabla}}_t+\frac{\Omega_s}{2}a_s\overrightarrow{\boldsymbol{\slashed{\nabla}}}, \nonumber \\
&& \bar{\Omega}=1+\frac{\bar{\Omega}_m}{2}a_tm_c+\frac{\bar{\Omega}_t}{2}a_t\overleftarrow{\slashed{\nabla}}_t+\frac{\bar{\Omega}_s}{2}a_s\overleftarrow{\boldsymbol{\slashed{\nabla}}}, 
\label{eq:A1}
\end{eqnarray}
with $\Omega_{m,\,t,\,s}$ (and $\bar{\Omega}_{m,\,t,\,s}$) being mass-dependent pure numbers, and where the covariant lattice derivatives $\nabla_{\mu}$ are defined as
\begin{equation}
\nabla_{\mu}\psi(x)=\frac{1}{2a_{\mu}}\big[U_{\mu}(x)\psi(x+\mu)-U_{-\mu}(x)\psi(x-\mu)\big]. \nonumber
\end{equation}

The application of this field redefinition to the anisotropic action
is discussed in detail in Ref.~\cite{Chen:2001hq}.  Here we will focus
on the improved quark-bilinear operators, given by
\begin{equation}
J^I=\bar{\psi}_c\Gamma\psi_c=\bar{\psi}\,\bar{\Omega}\,\Gamma\,\Omega\,\psi,
\label{eq:A2}
\end{equation}
which, after substitution the formulae from Eqn. \eqref{eq:A1} and requiring $\Omega_m=\bar{\Omega}_m$,  $\Omega_t=\bar{\Omega}_t$ and  $\Omega_s=\bar{\Omega}_s$ (see \cite{Chen:2001hq}) turns into 
\begin{align}
J^I&=(1+m_ca_t\Omega_m)J^U+\frac{1}{2}\Omega_ta_t[\bar{\psi}\Gamma\overrightarrow{\slashed{\nabla}}_t\psi-\bar{\psi}\overleftarrow{\slashed{\nabla}}_t\Gamma\psi]+\nonumber \\
&+\frac{1}{2}\Omega_sa_s[\bar{\psi}\Gamma\overrightarrow{\boldsymbol{\slashed{\nabla}}}_s\psi-\bar{\psi}\overleftarrow{\boldsymbol{\slashed{\nabla}}}_s\Gamma\psi],
\label{eq:A3}
\end{align}
where $J^U\equiv \bar{\psi}\Gamma\psi$ is the unimproved operator.

For the case of the axial-vector current we have
$\Gamma=\gamma_{\mu}\gamma_5$, and the improved axial-vector current 
current $A^I_{\mu}$ is given by
\begin{align}
  A_{\mu}^I&=(1+\Omega_ma_tm_c)A_{\mu}^U +\frac{\Omega_ta_t}{2}(\bar{\psi}\Gamma\overrightarrow{\slashed{\nabla}}_t\psi - \bar{\psi}\overleftarrow{\slashed{\nabla}}_t\Gamma\psi)+ \nonumber \\
  &+\frac{\Omega_sa_s}{2}(\bar{\psi}\Gamma\overrightarrow{\boldsymbol{\slashed{\nabla}}}_s\psi - \bar{\psi}\overleftarrow{\boldsymbol{\slashed{\nabla}}}_s\Gamma\psi) =\nonumber \\
  &=(1+\Omega_ma_tm_c)A_{\mu}^U +\frac{\Omega_ta_t}{2}(\bar{\psi}\gamma_{\mu}\gamma_5\gamma_4\overrightarrow{D}_4\psi - \nonumber \\
  &-\bar{\psi}\gamma_4\gamma_{\mu}\gamma_5\overleftarrow{D}_4\psi)+\frac{\Omega_sa_s}{2}(\bar{\psi}\gamma_{\mu}\gamma_5\gamma_j\overrightarrow{D}_j\psi - \nonumber \\
  &-\bar{\psi}\gamma_j\gamma_{\mu}\gamma_5\overleftarrow{D}_j\psi).
\label{eq:A4}
\end{align}
Using the relationship between the Euclidean gamma matrices and the Dirac matrices,
\begin{equation}
\gamma_{\mu}\gamma_{\nu}=\delta_{\mu\nu} + \sigma_{\mu\nu},
\label{eq:A5}
\end{equation}
where
\begin{equation}
\sigma_{\mu\nu} =  \frac{1}{2}[\gamma_{\mu},\ \gamma_{\nu}],
\end{equation}
Eqn.~\eqref{eq:A4} can be re-written as:
\begin{align}
A_{\mu}^I&=\big(1+\Omega_ma_tm_c\big)A_{\mu}^U -\nonumber \\
&-\frac{\Omega_ta_t}{2}\big(\bar{\psi}(\delta_{\mu4}+\sigma_{\mu4})\gamma_5\overrightarrow{D}_4\psi+\bar{\psi}(\delta_{4\mu}+\sigma_{4\mu})\gamma_5\overleftarrow{D}_4\psi \big) -\nonumber \\
&-\frac{\Omega_s a_s}{2}\big(\bar{\psi}(\delta_{\mu j}+\sigma_{\mu j})\gamma_5\overrightarrow{D}_j\psi+\bar{\psi}(\delta_{j\mu}+\sigma_{j\mu})\gamma_5\overleftarrow{D}_j\psi\big),
\label{eq:A6}
\end{align}
or
\begin{align}
A_{\mu}^I&=\big(1+\Omega_ma_tm_c\big)A_{\mu}^U+\nonumber \\
&+\frac{\Omega_ta_t}{2}\big(-\delta_{\mu4}\partial_4\bar{\psi}\gamma_5\psi - \sigma_{\mu4}\bar{\psi}\gamma_5\overrightarrow{D}_4\psi-\sigma_{4\mu}\bar{\psi}\gamma_5\overleftarrow{D}_4\psi\big)+\nonumber \\
&+\frac{\Omega_sa_s}{2}\big(-\delta_{\mu j}\partial_j\bar{\psi}\gamma_5\psi -\sigma_{\mu j}\bar{\psi}\gamma_5\overrightarrow{D}_j\psi-\sigma_{j\mu}\bar{\psi}\gamma_5\overleftarrow{D}_j\psi\big)
\label{eq:A7}
\end{align}
To simplify this expression, we make use of the equations of motion
which (to the lowest order) are written as:
\begin{eqnarray}
&& (m_0+\nu_t\overrightarrow{\slashed{D}}_t+\nu_s\overrightarrow{\slashed{D}}_s)\psi=0,\label{eq:A8}  \\ 
&& \bar{\psi}(m_0-\nu_t\overleftarrow{\slashed{D}}_t-\nu_s\overleftarrow{\slashed{D}}_s)=0. \label{eq:A9}
\end{eqnarray}
From the first equation:
\begin{equation}
m_0\gamma_{\rho}\psi+\nu_t\gamma_{\rho}\gamma_4\overrightarrow{D}_4\psi+\nu_s\gamma_{\rho}\gamma_j\overrightarrow{D}_j\psi=0,
\label{eq:A10}
\end{equation}
and therefore
\begin{equation}
\nu_s\sigma_{\rho j}\overrightarrow{D}_j\psi + \nu_t\sigma_{\rho 4}\overrightarrow{D}_4\psi=-m_0\gamma_{\rho}\psi-\nu_t\delta_{\rho 4}\overrightarrow{D}_4\psi-\nu_s\delta_{\rho j}\overrightarrow{D}_j\psi.
\label{eq:A11}
\end{equation}
Similarly, from Eqn. \eqref{eq:A9} we get:
\begin{equation}
m_0\bar{\psi}\gamma_{\rho}-\nu_t\bar{\psi}\gamma_4\gamma_{\rho}\overleftarrow{D}_4-\nu_s\bar{\psi}\gamma_j\gamma_{\rho}\overleftarrow{D}_j=0,
\label{eq:A12}
\end{equation}
and
\begin{equation}
\nu_t\bar{\psi}\sigma_{4\rho}\overleftarrow{D}_4+\nu_s\bar{\psi}\sigma_{j\rho}\overleftarrow{D}_j=m_0\bar{\psi}\gamma_{\rho}-\nu_t\bar{\psi}\delta_{4\rho}\overleftarrow{D}_4-\nu_s\bar{\psi}\delta_{j\rho}\overleftarrow{D}_j.
\label{eq:A13}
\end{equation}
Here we consider the temporal component of the axial-vector current
($\mu=4$), so Eqn.~\eqref{eq:A7} becomes
\begin{align}
A_4^I&=\big(1+\Omega_ma_tm_c\big)A_4^U-\frac{\Omega_ta_t}{2}\delta_{\rho4}\partial_4\bar{\psi}\gamma_5\psi-\nonumber \\
&-\frac{\Omega_sa_s}{2}\big(\sigma_{4 j}\bar{\psi}\gamma_5\overrightarrow{D}_j\psi+\sigma_{j4}\bar{\psi}\gamma_5\overleftarrow{D}_j\psi\big)
\label{eq:A14}
\end{align}
and, after applying Eqns.~\eqref{eq:A11} and \eqref{eq:A13}, we obtain:
\begin{equation}
A_4^I=(1+\Omega_ma_tm_c)A_4^U+\frac{a_t}{2}\big(\Omega_s\frac{a_s}{a_t}\frac{\nu_t}{\nu_s}-\Omega_t\big)\partial_4\bar{\psi}\gamma_5\psi.
\label{eq:A15}
\end{equation}
We choose the case with $\nu_t=1$ (so-called ``$\nu_s$-tuning"), where $\nu_s$
is tuned via the dispersion relation between meson energy and
momentum, yielding
\begin{equation}
\nu_s=\frac{1+\frac{1}{2}a_tm_c}{1+\frac{1}{2}a_sm_c}.
\label{eq:A16}
\end{equation}
The parameters $\Omega_t$ and $\Omega_s$ are set as in Ref.~\cite{Chen:2001hq}:
\begin{equation}
\Omega_s=-\frac{1}{2}\left(\frac{1+\frac{1}{2}a_tm_c}{1+\frac{1}{2}a_sm_c}\right), \, \, \, \Omega_t=-\frac{1}{2}.
\label{eq:A17}
\end{equation}
The value for the anisotropy parameter in our calculations is
$\xi=\frac{a_s}{a_t}\approx 3.5$, so the final expression for the time
component of the improved axial-vector current takes the form:
\begin{equation}
A_4^I=(1+\Omega_ma_tm_c)A_4^U-0.625\,a_t\partial_4\bar{\psi}\gamma_5\psi,
\label{eq:A18}
\end{equation}
or, up to leading order in $a$,
\begin{equation}
A_4^I=(1+\Omega_ma_tm_c)\left [A_4^U-\frac{1}{4}(\xi-1)a_t\partial_4P\right ].
\label{eq:A19}
\end{equation}
\end{appendix}
\bibliography{old_paper_mayber} {}
\end{document}